\documentclass[
reprint,
superscriptaddress,
amsmath,
amssymb,
aps,
longbibliography,
pra,
showpacs,
floatfix
]{revtex4-1}

\usepackage[mathlines]{lineno}
\usepackage{graphicx}
\usepackage[margin=1in]{geometry} 
\usepackage{graphicx}% Include figure files
\usepackage[table]{xcolor}% colouring text also in tables

\usepackage{tabularx}% better tables
\usepackage{multirow}
\usepackage{dcolumn} 
\newcolumntype{d}[1]{D{.}{.}{#1}}
\newcolumntype{L}[1]{>{\raggedright\arraybackslash}p{#1}}
\newcolumntype{C}[1]{>{\centering\arraybackslash}p{#1}}
\newcolumntype{R}[1]{>{\raggedleft\arraybackslash}p{#1}}
\usepackage{booktabs}

\usepackage{comment}% comment multiple lines
\usepackage{enumitem}% use [noitemsep] in itemize for smaller whitespace

\usepackage[hidelinks]{hyperref}
\definecolor{color1}{rgb}{0,0.25,0.70}
\hypersetup{colorlinks=true, 
    linkcolor={color1},
    citecolor={color1}, 
    urlcolor={color1}
}

\usepackage[]{xcolor}
\usepackage{soul}
\usepackage{cancel,siunitx}

\renewcommand{\epsilon}{\varepsilon}

\usepackage{physics}
\usepackage{bm}

\graphicspath{ {./figures/} }

\begin{document}

\preprint{APS/123-QED}

\title{Chemical reactions in imperfect cavities: enhancement, suppression, and resonance}

\author{John P. Philbin$^{\P}$}
 \email{jphilbin@g.harvard.edu}
 \affiliation{Harvard John A. Paulson School of Engineering and Applied Sciences, Harvard University, Cambridge, MA 02138, USA%
}

\author{Yu Wang$^{\P}$}
 \email{wangyu19@westlake.edu}
 \affiliation{Department of Chemistry, School of Science, Westlake University, Hangzhou 310024 Zhejiang, China}
 \affiliation{Institute of Natural Sciences, Westlake Institute for Advanced Study, Hangzhou 310024 Zhejiang, China
}

\thanks{Denotes equal contribution}

\author{Prineha Narang}
 \email{prineha@seas.harvard.edu}
 \affiliation{Harvard John A. Paulson School of Engineering and Applied Sciences, Harvard University, Cambridge, MA 02138, USA%
}

\author{Wenjie Dou}
 \email{douwenjie@westlake.edu}
 \affiliation{Department of Chemistry, School of Science, Westlake University, Hangzhou 310024 Zhejiang, China}
 \affiliation{Department of Physics, School of Science, Westlake University, Hangzhou 310024 Zhejiang, China}
 \affiliation{Institute of Natural Sciences, Westlake Institute for Advanced Study, Hangzhou 310024 Zhejiang, China
}

\date{\today}

\begin{abstract}
The use of optical cavities to control chemical reactions has been of great interest recently, following demonstrations of enhancement, suppression, and negligible effects on chemical reaction rates depending on the specific reaction and cavity frequency. In this work, we study the reaction rate inside imperfect cavities, where we introduce a broadening parameter in the spectral density to mimic Fabry-P\'erot cavities. We investigate cavity modifications to reaction rates using non-Markovian Langevin dynamics with frictional and random forces to account for the presence of imperfect optical cavities. We demonstrate that in the regime of weak solvent and cavity friction, the cavity can enhance chemical reaction rates. On the other hand, in the high friction regime, cavities can suppress chemical reactions. Furthermore, we find that the broadening of the cavity spectral density gives rise to blue shifts of the resonance conditions and, surprisingly, increases the sharpness of the resonance effect. 
\end{abstract}

\maketitle

\section{Introduction} \label{sec:intro}

The concept of selectively controlling the rate of a chemical reaction by shining light in resonance with a specific molecular bond has been explored for many decades. Two drawbacks of this scheme are that it requires light energy as an input and, arguably more importantly, molecular bonds often quickly transfer the vibrational energy in the mode of interest to other vibrations and degrees of freedom. A possible way to circumvent these two obstacles is to tune the frequencies of an optical cavity to selectively alter the ground state chemical reactivity of molecules, which was demonstrated recently experimentally.\cite{Thomas2016,ThomasScience2019,Nagarajan2021} This research avenue has attracted many experimental,\cite{Lather2019,Wiesehan2021,Imperatore2021,Mandal2022} computational,\cite{Ruggenthaler2014,Flick2017,Flick2018,FregoniNatureCommun2018,Flick2020,Haugland2020,Deprince2021,Schafer2021,Pavosevic2022} and theoretical\cite{campos2019resonant,Li2020b,Li2021a,Yang2021} investigations in recent years as which reactions can be manipulated by optical cavities and the underlying mechanism of how cavities can modify chemical reactivity are still an open questions.\cite{Kena-Cohen2019,hertzog2019strong,Simpkins2021}

In this work, we study the reaction rate inside an optical cavity by developing a non-Markovian dynamical model. The effect of the cavity mode on the reaction mode dynamics is incorporated into friction and random forces.\cite{PhysRevLett.119.046001,dou2018perspective} This model allows us to continuously model Fabry-P\'erot cavities from their ideal perfect single mode cavity limit to more realistic lossy, imperfect cavities.\cite{Wang2021} In Kramer's theory dealing with Ohmic friction, the rate of escape of a particle over a potential barrier increases linearly with increasing friction in the underdamped limit and decreases inversely with the friction strength in the strong damping limit.\cite{kramers1940brownian} Grote and Hynes proceeded to solve the Kramer's rate problem in the presence of memory friction, in the regime of moderate to strong friction.\cite{grote1980stable} The continuum limit version of Kramer's theory that cover the whole range of friction is known as Pollak, Grabert, and H\"anggi (PGH) theory.\cite{pollak1989theory} To cover the whole range of friction, our numerical and analytical analysis show that optical cavities can both enhance or suppress chemical reaction rates, depending on the magnitude of the solvent friction as well as the cavity-molecule coupling strength. Interestingly, increasing the cavity-molecule coupling strength gives rise to a blue shift of the cavity frequency that has the greatest impact on the reaction rate (i.e. blue shift of the resonance condition) whereas making the cavity more imperfect by increasing the broadening parameter of the cavity spectral density results in a blue shift of the resonance condition. 

The manuscript is organized as follows. In Sec.~\ref{sec:theory}, we introduce our model and derive the Langevin dynamics for the molecules in the presence of a Markovian phonon bath and non-Markovian cavity mode. In Sec.~\ref{sec:results}, we perform both numerical and analytical analysis to obtain cavity-modified chemical reaction rates within our model. Lastly, we conclude in Sec.~\ref{sec:conclusion}. 

\begin{figure}
    \centering
    \includegraphics[scale=0.25]{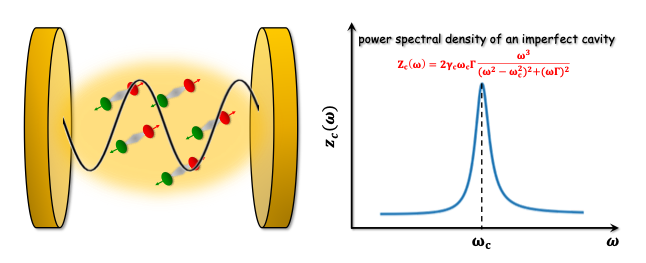}
    \caption{Schematic representation of light-matter vibrational strong coupling in an imperfect cavity and the spectra density of the cavity.}
    \label{fig:schematic}
\end{figure}

\section{Theory} \label{sec:theory}

Here, we consider the reaction mode to be a one-dimension double well potential energy surface (PES). The reaction mode couples to a set of harmonic oscillators corresponding to the solvent environment as well as a set of harmonic oscillators corresponding to the optical cavity. The total Hamiltonian that we utilize is given by the Pauli-Fierz nonrelativistic quantum electrodynamics (QED) Hamiltonian in the dipole gauge and in the long-wavelength limit:\cite{RokajJPhysB2018,Schafer2020} 
\begin{align}
\nonumber H = &~\frac{p^2}{2m} + V(x) + \frac12\sum_j m_j [ \dot q_j^2 + \omega_j^2 ( q_j + c_j x)^2  ] \\ & + \frac12 \sum_k [ P_k^2 +  \tilde \omega_k^2 ( Q_j + \mu_k x)^2  ] .
\end{align}
Here, $x$ and $p$ are the coordinates for the reaction mode, $q_j$ and $\dot q_j$ are the coordinates for the other vibrational modes that will compose the Markovian bath, $Q_k$ and $P_k$ are the coordinates for the cavity modes that will compose the non-Markovian bath, $c_j$ and $\mu_k$ are coupling strengths of the reaction coordinates to the vibrational bath modes and cavity modes. The double well PES for the reaction mode is given by \cite{dou2015frictional}: 
\begin{equation}
V(x) = \frac12m \omega_0^2x^2-a\cdot \log(1+e^{-(\sqrt{2}gx+E_d)/a}),
\end{equation}
where $\omega_0=0.003$,  $a=0.02$, $g=0.02$, $E_d=\frac{40}{3}$ are the parameters used in our calculations below and we set $\hbar=1$ throughout this work. In the semi-classical limit, we can write down the equations of motion for the reaction mode as: 
\begin{align}
\label{eq:eom_x} m \ddot x = & - \frac{dV}{dx} - \sum_j m_j \omega_j^2 (c_j q_j +c_j^2 x) \\ \nonumber & -\sum_k  \tilde \omega_k^2 (\mu_k Q_k +\mu_k^2 x) .
\end{align}
Similarly, we can also write down the equation of motions for the vibrational bath modes and cavity modes 
\begin{align}
m_j \ddot q_j & = - m_j \omega_j^2  q_j - m_j \omega_j^2 c_j x \\
\ddot Q_k & = - \tilde \omega_k^2  Q_k - \tilde \omega_k^2 \mu_k x .
\end{align}
Because the above equations are linear, they can be solved explicitly: \cite{Nitzan2006} 
\begin{align}
\label{eq:q_j(t)} q_j(t) = &~q_{j0} \cos(\omega_j t) + \frac{\dot q_{j0}}{\omega_j} \sin (\omega_j t) - c_j x(t) \\ \nonumber & + c_j \int_0^t d\tau \cos(\omega_j(t-\tau)) \dot x(\tau) ,    
\end{align}
where $q_{j0}$ and $\dot q_{j0}$ are the initial position and momenta for the solvent environment. We can solve for $P_k, Q_k$ in a similar manner,
\begin{align}
\label{eq:Q_k(t)} Q_k(t) = &~Q_{k0} \cos(\tilde \omega_k t) + \frac{\dot Q_{k0}}{\tilde \omega_k} \sin (\tilde \omega_k t) - \mu_k x(t) \\ \nonumber & + \mu_k \int_0^t d\tau \cos(\tilde \omega_k (t-\tau)) \dot x(\tau) .
\end{align}

If we plug in Eq.~(\ref{eq:q_j(t)}) and Eq.~(\ref{eq:Q_k(t)}) into the equation of motion for $x$ given by Eq.~(\ref{eq:eom_x}), we find a generalized Langevin equation: 
\begin{align}
\nonumber m \ddot x = & - \frac{dV}{dx}  - \int_0^t d\tau (Z_p (t-\tau) + Z_c (t-\tau) ) \dot x(\tau) \\ & + R_p (t) + R_c (t)
\end{align}
where
\begin{align}
Z_p (t) & = \sum_j m_j \omega_j^2 c_j^2 \cos(\omega_j t)  \\
Z_c (t) & = \sum_k \tilde \omega_k^2 \mu_k^2 \cos(\tilde \omega_k t) \\
R_p(t) & = - \sum_j m_j \omega_j^2 c_j (q_{j0} \cos(\omega_j t) + \frac{\dot q_{j0}}{\omega_j} \sin (\omega_j t)) \\
R_c (t) & = - \sum_k \tilde \omega_k^2 \mu_k (Q_{k0} \cos(\tilde \omega_k t) + \frac{\dot Q_{k0}}{\tilde \omega_k} \sin (\tilde \omega_k t) ).
\end{align}
Here $R_p(t)$ and $R_c(t)$ are random forces from the phonon bath and cavity respectively, and $Z_c(t)$ and $Z_p(t)$ are the corresponding friction kernels. Below we consider the case that friction and random forces from the phonon environments are Markovian, such that the Langevin equation can be simplified as  
\begin{align}\label{dy1}
m \ddot x = &- \frac{dV}{dx}  - \int_0^t d\tau Z_c (t-\tau) \dot x(\tau) + R_c (t) \\ \nonumber & + \gamma_p \dot x(t) + R_p (t)
\end{align}

We now turn our attention to the frictional force from the cavity. We consider the spectral density with broadening parameters in an imperfect cavity.\cite{Wang2021} In particular, the spectral density is taken to the form of a Cauchy distribution to mimic a Fabry-P\'erot cavity:
\begin{eqnarray}\label{Zw}
Z_c(\omega)=2\gamma_c\omega_c\Gamma\frac{\omega^3}{(\omega^2-\omega_c^2)^2+(\omega\Gamma)^2}.
\end{eqnarray}
In the above equation (Eq.~\ref{Zw}), $\gamma_c$ quantifies the coupling strength of the reaction coordinate to the cavity, $\omega_c$ corresponding to the the center frequency where the cavity density of states is largest (and is the frequency of the one cavity mode in the perfect cavity scenario), and $\Gamma$ is the broadening parameter that determines the extent to which the cavity is imperfect (i.e. lossy). As $\Gamma \rightarrow 0$, $Z_c(\omega)$ approaches a $\delta$-function, reflecting a perfect cavity. 

\begin{figure}
    \centering
    \includegraphics[scale=0.35]{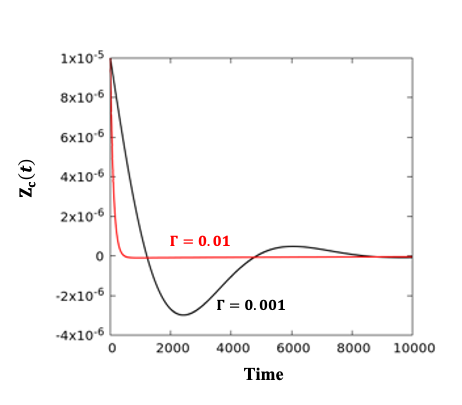}
    \caption{The memory kernel $Z_c(t)$ as a function of time with a small $\Gamma$ (black line) and a large $\Gamma$ (red line). The other model parameters used here are: $\gamma_c=0.01$, $\omega_c=0.001$.}
    \label{fig:kernel}
\end{figure}

The memory kernel is then given by
\begin{eqnarray}
Z_c(t) = \frac{2}{\pi} \int_0^\infty \frac{Z_c(\omega)}{\omega} \cos{(\omega t)} d\omega
\end{eqnarray}
The solution cavity memory kernel in the time domain has two solutions, depending on the cavity center frequency and broadening. When $\omega_c>\Gamma/2$, the cavity memory kernel is 
\begin{align}
Z_c (t)=2\gamma_c\omega_c e^{-\frac{\Gamma t}{2}}\left(\frac{ \cos(\omega_1 t)}{2}-\frac{\Gamma \sin(\omega_1 t)}{4\omega_1}\right)
\end{align}
where we defined $\omega_1=\sqrt{\omega_c^2-\Gamma^2/4}$. And when $\omega_c<\Gamma/2$, the cavity memory kernel is
\begin{align}
Z_c (t)=2\gamma_c\omega_c e^{-\frac{\Gamma t}{2}}\left(\frac{ \cosh(\omega_1 t)}{2}-\frac{\Gamma \sinh(\omega_1 t)}{4\omega_1}\right).
\end{align}
with $\omega_1=\sqrt{\Gamma^2/4-\omega_c^2}$. As shown in Fig.~\ref{fig:kernel}, the cavity memory kernel in the time domain converges toward a $\delta$-function as $\Gamma$ increases. Whereas, when $\Gamma$ approaches to 0, the memory kernel will oscillate as a cosine function. The cavity random force $R_c(t)$ is related to the memory kernel $Z_c(t)$ via the fluctuation–dissipation theorem:
\begin{align}\label{corr_Rc}
\langle R_c(0)R_c (t)\rangle=m k_B T Z_c(t).    
\end{align}
Note that the time correlation function of the random force is non-Markovian. To generate such non-Markovian random force, we propagate the following equations of motion: 
\begin{align}\label{dy2}
 \dot Y=\sqrt{\frac{\omega_c}{\Gamma}}\omega_c R_c
\end{align}
\begin{align}\label{dy3}
 \dot R_c =\sqrt{\frac{\omega_c}{\Gamma}}(-\omega_c Y - \Gamma R_c + \Gamma \xi(t)).
\end{align}
Here, $\xi(t)$ is a Markovian random variable from a Gaussian distribution with a standard deviation of $\sigma=\sqrt{2m \gamma_ck_BT/ dt}$, where $dt$ is the time step interval. $Y$ is an auxiliary variable. We can show that, in the long time limit, the above equations generate $R_c$ that satisfies the correlation function in Eq.~(\ref{corr_Rc}). 

As for the solvent (i.e. vibrational) bath, the random force $R_p(t)$ is Markovian ($\langle R_p(0)R_p (t)\rangle= 2 m k_B T \gamma_p \delta(t)$), which is set to be a Gaussian random variable with a standard deviation of $\sigma_p=\sqrt{2m \gamma_pk_BT/ dt)}$. We use $4^{\text{th}}$-order Runge-Kutta to integrate Eqs.~(\ref{dy1}), (\ref{dy2}) and (\ref{dy3}). Lastly, unless stated otherwise, we perform thermal averages over $10,000$ trajectories in our non-Markovian Langevin dynamics simulations.

\section{Results and Discussion} \label{sec:results}

\begin{figure}
    \centering
    \includegraphics[scale=0.3]{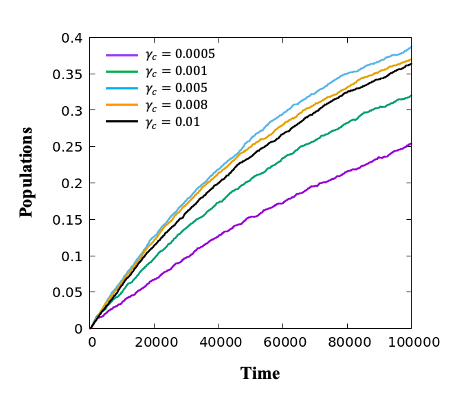}
    \caption{Product populations as a function of time with different cavity friction $\gamma_c$, calculated by non-Markovian Langevin dynamics. The other model parameters used here are: $\omega_0=0.003$, $\omega_b=0.007$, barrier height $E_b=0.01947$, $\gamma_p=0.00001$, $\omega_c=0.005$, $k_{\text{B}}T=0.005$, $\Gamma=0.01$.}
    \label{fig:rate}
\end{figure}

We first investigate cavity modified reaction rate in the weak solvent friction regime, namely when $\gamma_p$ is small. We calculate the reaction rate by initializing all trajectories in the reactant well with a Boltzmann distribution and monitoring the number of trajectories that end up in the product well as a function of time by numerically solving the non-Markovian Langevin dynamics, as illustrated in Fig.~\ref{fig:rate}. We find that, with a fixed photon frequency $\omega_c=0.005$, a turnover occurs in the chemical reaction rate as a function of the cavity friction $\gamma_c$. Specifically, we find that initially increasing the cavity friction results increases the reaction rate but, eventually, increasing the cavity friction leads to a decrease in the reaction rate. This turnover is the well-known Kramer's turnover which occurs in the overdamped regime.\cite{Hanggi1990} Therefore, with small solvent friction, the Kramer turnover effect can be observed here as a function of the cavity friction.

\begin{figure}
    \centering
    \includegraphics[scale=0.3]{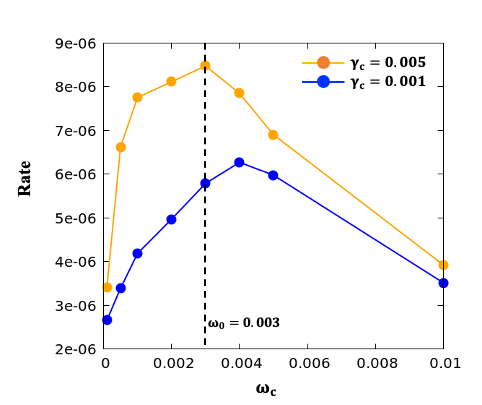}
    \caption{Cavity-modified reaction rate as a function of the cavity photon frequency $\omega_c$ at two different cavity friction values, $\gamma_c=0.001, 0.005$, calculated by non-Markovian Langevin dynamics. The other model parameters used here are: $\omega_0=0.003$, $\omega_b=0.007$, barrier height $E_b=0.01947$, $\gamma_p=0.00001$, $k_{\text{B}}T=0.005$, $\Gamma=0.01$.}
    \label{fig:fric_p_small}
\end{figure}

We then proceed to investigate the reaction rate as a function of the cavity frequency, where the rate is extracted by exponentially fitting the product population as a function of time. In the limit of weak solvent friction, we see that the cavity can enhance reaction rates ($\omega_c \approx \omega_0 $) (Fig.~\ref{fig:fric_p_small}). The maximum cavity-modified reaction rate enhancement occurs at a cavity frequency around the vibrational frequency of the reactant well ($\omega_0=0.003$, Fig.~\ref{fig:fric_p_small}). Upon increasing the cavity friction $\gamma_c$ from $0.001$ to $0.005$, the cavity enhances the reaction even more as the total friction remains in the underdamped regime. Further increasing the cavity friction will eventually results in suppressing chemical reactions as shown in Fig.~\ref{fig:rate} (overdamped regime). 

We now turn our focus to the strong solvent friction regime, namely when $\gamma_p$ is large. In this regime, the reaction rates are monotonically suppressed upon increasing the cavity friction, which is in agreement with Grote-Hynes (GH) theory.\cite{grote1980stable,Li2021a,Mandal2022} In this limit, the reaction rate is given by 
\begin{eqnarray}
k = \kappa_{\text{GH}} k_{\text{TST}}
\end{eqnarray}
Here $k_{\text{TST}}$ is the transition state theory rate. $\kappa_{\text{GH}}$ is the Grote-Hynes transmission coefficient. We can employ an analytical approach for calculating $\kappa_{\text{GH}}$ to understand how cavities can modify chemical reactions in this regime as a function of the cavity frequency, $\omega_c$. Particularly, the GH coefficient is given by
\begin{eqnarray}
\lambda^2=\omega_b^2-\lambda\cdot(\tilde{\xi}_c(\lambda)+\tilde{\xi}_p(\lambda)), 
\end{eqnarray}
where $\tilde{\xi}_c(\lambda)$ and $\tilde{\xi}_p(\lambda)$ are the Laplace transformed cavity photon friction and solvent vibrational bath friction, respectively. Using the spectral density $Z_c(\omega)$, one can write $\tilde{\xi}_c(\lambda)$ as:
\begin{align}
\tilde{\xi}_c(\lambda)=\frac{2}{\pi}\int_0^\infty\frac{Z_c(\omega)}{\omega}\frac{\lambda}{\lambda^2+\omega^2}d\omega.
\end{align}
For solvent bath friction, we have $\tilde{\xi}_p = \gamma_p$. GH coefficient is then obtained from $\kappa_{\text{GH}} = \lambda /\omega_b$.

\begin{figure}
    \centering
    \includegraphics[scale=0.3]{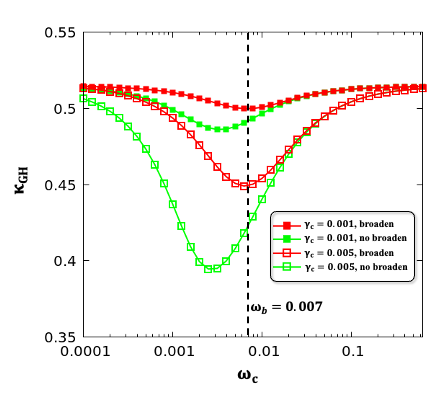}
    \caption{Cavity-modified transmission coefficient $\kappa_{\text{GH}}$ for the reaction rate as a function of the cavity photon frequency $\omega_c$ at different cavity frictions $\gamma_c=0.001, 0.005$. The other model parameters used here are: $\omega_0=0.003$, $\omega_b=0.007$, barrier height $E_b=0.01947$, $\gamma_p=0.01$, $k_{\text{B}}T=0.005$, for the green lines, $\Gamma=0.0001$, for the red lines, $\Gamma=0.01$.}
    \label{fig:fricN_large}
\end{figure}

In Fig.~\ref{fig:fricN_large}, we plot $\kappa_{\text{GH}}$ as a function of $\omega_c$ analytically. The two green lines show reaction rates for two perfect cavities with different cavity frictions ($\gamma_c$) in comparison to two red lines which show reaction rates for imperfect cavities (i.e. cavities with broadened spectral densities) with all other parameters held constant. In agreement with recent reports,\cite{Li2021a,Mandal2022} when the cavity friction ($\gamma_c$) is small, the minimum of $\kappa_{\text{GH}}$ appears at a cavity frequency close to the barrier frequency, $\omega_b=0.007$. Upon increasing $\gamma_c$, the cavity frequency corresponding to the minimum of $\kappa_{\text{GH}}$ is shifted to lower energies (i.e. the resonance condition is red shifted). A noteworthy finding reported here for the first time is that effect of imperfect cavities on the resonance conditions. Fig.~\ref{fig:fricN_large} shows that the cavity frequency with greatest impact on the reaction rates is shifted to higher energies (i.e. blue shifted). This is shown clearly by the maximum rate suppression occurring at higher cavity frequencies for the imperfect cavities (red lines) compared to the corresponding perfect cavities (green lines). Additionally, the resonance condition surprisingly sharpens in the case of a broaden cavity, especially on the low frequency side of the maximum rate suppression.

To understand the red and blue shifts, we examine the GH theory again. 
To simplify the analysis here, we eliminate the friction from phonon bath. In the limit of $\Gamma = 0$ (perfect cavity), the GH equation becomes 
\begin{eqnarray}\label{GH1}
\lambda^2=\omega_b^2- 2 \gamma_c\omega_c \frac{\lambda^2}{\lambda^2+\omega_c^2}.
\end{eqnarray}
We can solve for $\omega_c$ where we reach minimum $\kappa_{\text{GH}}$ by finding the solution $\partial \lambda /\partial \omega_c = 0$ (see the Appendix): 
\begin{eqnarray}\label{eq:wc_min2}
\omega_c=-\frac{\gamma_c}{2}+\sqrt{\frac{\gamma_c^2}{4}+\omega_b^2} \approx \omega_b - \frac{\gamma_c}{2}
\end{eqnarray}
Here, we have expanded the above result to the first order in $\gamma_c$. Clearly, the coupling strength $\gamma_c$ will introduce a red shift for the resonance condition $\omega_c = \omega_b$.  

For the imperfect cavity with non-zero $\Gamma$, we can introduce a factor $Q$ to approximate the following integral
\begin{align}
\frac{2}{\pi}\int_0^\infty \frac{\Gamma\omega^2}{(\omega^2-\omega_c^2)^2+(\omega\Gamma)^2} \frac{\lambda}{\lambda^2+\omega^2}d\omega = \frac{\lambda}{\lambda^2+(Q\omega_c)^2}.
\end{align}
In the limit $\Gamma=0$, we have $Q=1$. Such that we recover the results for the perfect cavity. The non-zero $\Gamma$ will lead to $Q<1$ for an imperfect cavity. The minimum $\kappa_{\text{GH}}$ is then located at 
\begin{eqnarray}\label{eq:wc_min2}
\omega_c=\frac{1}{Q^2}\left(-\frac{\gamma_c}{2}+\sqrt{\frac{\gamma_c^2}{4}+\omega_b^2}\right)
\end{eqnarray}
Since $Q<1$, the broadening will give rise to blue shift.

\begin{figure}
    \centering
    \includegraphics[scale=0.25]{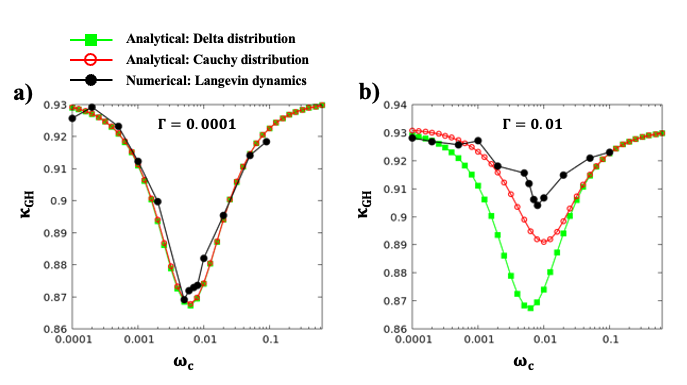}
    \caption{Cavity-modified transmission coefficient $\kappa_{\text{GH}}$ for the reaction rate as a function of photon frequency $\omega_c$ for (a) a perfect cavity ($\Gamma=0.0001$) and (b) an imperfect cavity ($\Gamma=0.01$). The other model parameters used here are: $\omega_0=0.003$, $\omega_b=0.007$, barrier height $E_b=0.01947$, $k_{\text{B}}T=0.005$, $\gamma_p=0.001$, $\gamma_c=0.001$.}
    \label{fig:kappa}
\end{figure}

In Fig.~\ref{fig:kappa}, we further verify the effects of broadening on the transmission coefficient $\kappa_{\text{GH}}$ as a function of the cavity frequency $\omega_c$ using numerical and analytic methods. Specifically, the green lines and red lines utilize the analytic expressions for a perfect and imperfect cavity, respectively and the black lines are the numerical results obtained from the full non-Markovian Langevin dynamics with the same $\Gamma$ as the red lines. Fig.~\ref{fig:kappa}a shows that when $\Gamma$ is small, there is no broadening effect in the spectral density of the cavity mode such that these three methods predict the same $\kappa_{\text{GH}}$ for all values of $\omega_c$. In particular, $\kappa_{\text{GH}}$ exhibits a minimum when the cavity frequency is close to the barrier frequency ($\omega_b=0.007$). On the other hand, for the relatively larger $\Gamma$ value shown in Fig.~\ref{fig:kappa}b, the broadening effect of the cavity spectral density leads to a blue shift in the frequency at which the cavity has the largest impact on the reaction rate, and the width of the resonance narrows. This observation is in agreement with our analytical analysis above. Furthermore, the broadening effects tend to reduce the efficiency of the cavity mode on the reaction rate. With a very large $\Gamma$, we expect that the cavity will not affect the chemical reaction at all.

\section{Conclusions} \label{sec:conclusion}

We developed a non-Markovian friction and random force model to study how imperfect optical cavities can modify chemical reaction rates. We found that in the small solvent friction regime, cavities can enhance chemical reaction rates while in the large solvent friction regime cavities suppress reactions. We also reported that imperfect cavities have resonance conditions that are both shifted to higher frequencies and sharper relative to otherwise identical perfect cavities. 

After completing this work, we noticed that two similar results were reported in the literature.\cite{Sun2022,Lindoy2022} Our work is in agreement with these findings and differentiates itself from those works as we investigated the effect of imperfect cavities, both numerically and analytically, on chemical reaction rates while the other reports focused on perfect cavities.

\begin{acknowledgments}
Y.W and W.D. acknowledge support from Westlake University. 
J.P.P. acknowledges support from the Harvard University Center for the Environment.
P.N. acknowledges support as a Moore Inventor Fellow through Grant No. GBMF8048 and gratefully acknowledges support from the Gordon and Betty Moore Foundation as well as support from a NSF CAREER Award under Grant No. NSF-ECCS-1944085.
\end{acknowledgments}

\appendix*
\section{}
As stated in the main text, to simplify the analysis in the large friction regime, we eliminate the friction from phonon bath. In the limit of $\Gamma = 0$ (perfect cavity), we have 
\begin{eqnarray}\label{eq:GH}
\lambda^2=\omega_b^2- 2 \gamma_c\omega_c \frac{\lambda^2}{\lambda^2+\omega_c^2}.
\end{eqnarray}
When $\frac{\partial\lambda}{\partial\omega_c}=0$ we have a relation between $\lambda_{min}$ and $\omega_c$ as  
\begin{eqnarray}\label{eq:lambda_min}
\lambda_{min}^2=\frac{\omega_b^2}{1+\frac{\gamma_c}{\omega_c}}
\end{eqnarray}
Substituting Eq. \ref{eq:lambda_min} into Eq. \ref{eq:GH} we could get a relationship as follows
\begin{eqnarray}\label{eq:lambda=wc}
\lambda_{min}=\omega_c
\end{eqnarray}
Substituting Eq. \ref{eq:lambda=wc}
into Eq. \ref{eq:lambda_min} obtains a function relating to the $\omega_c$ that corresponding to the $\lambda_{min}$ (i.e. the minimum of $\kappa_{\text{GH}}$)
\begin{eqnarray}\label{eq:wc_min1}
\omega_c^2+\gamma_c\omega_c-\omega_b^2=0
\end{eqnarray}
When solving the above equation we get
\begin{eqnarray}\label{eq:wc_min2}
\omega_c=-\frac{\gamma_c}{2}+\sqrt{\frac{\gamma_c^2}{4}+\omega_b^2}
\end{eqnarray}
It is evidently that when increasing the $\gamma_c$, the $\omega_c$ corresponding to the minimum $\kappa_{\text{GH}}$ will red shift.

For the imperfect cavity, when introducing a $Q$ factor, the GH equation becomes 
\begin{eqnarray}\label{eq:GH_Q}
\lambda^2=\omega_b^2- 2 \gamma_c\omega_c \frac{\lambda^2}{\lambda^2+ (Q\omega_c)^2}.
\end{eqnarray}
When $\frac{\partial\lambda}{\partial\omega_c}=0$ we have a relation between $\lambda_{min}$ and $\omega_c$ as  
\begin{eqnarray}\label{eq:lambda_min2}
\lambda_{min}^2=\frac{\omega_b^2}{1+\frac{\gamma_c}{Q^2\omega_c}}
\end{eqnarray}
Substituting Eq. \ref{eq:lambda_min2} into Eq. \ref{eq:GH_Q} we could obtain
\begin{eqnarray}\label{eq:lambda=wc_Q}
\lambda_{min}=Q\omega_c
\end{eqnarray}
Substituting Eq. \ref{eq:lambda=wc_Q} into Eq. \ref{eq:lambda_min2} gets a function relating to the $\omega_c$ that corresponding to the $\lambda_{min}$ in the imperfect cavity 
\begin{eqnarray}\label{eq:wc_min1}
Q^2\omega_c^2+{\gamma_c}\omega_c-\omega_b^2=0
\end{eqnarray}
When solving the above equation we get
\begin{eqnarray}\label{eq:wc_min2}
\omega_c=\frac{1}{Q^2}\left(-\frac{\gamma_c}{2}+\sqrt{\frac{\gamma_c^2}{4}+\omega_b^2}\right)
\end{eqnarray}
Note that $Q<1$. The resonance $\omega_c$ will blue shift in this imperfect cavity.

\bibliography{cavity} % produces the bibliography via BibTeX

\end{document}